\documentclass[intlimits,twoside,a4paper]{article}

\usepackage{graphicx}
\usepackage[cp1251]{inputenc}
\usepackage{amssymb}
\usepackage{amsmath}
\usepackage[usenames]{color}

\usepackage{epsfig}

\usepackage[eqsecnum]{cmpj3}

\issue{2019}{22}{3}{33601}
\doinumber{10.5488/CMP.22.33601}

\title{Phase diagrams and  critical behaviours of the mixed spin-5/2 and spin-7/2 Ising system}
\author[M. Karimou \textsl{et al.}]{M. Karimou\refaddr{label1,label2}\footnote{Corresponding author, E-mail: mounirou.karimou@yahoo.fr.}\,, R.A. Yessoufou\refaddr{label2,label3}, G. Dimitri Ngantso\refaddr{label4}, F. Hontinfinde\refaddr{label2,label3}, E. Albayrak\refaddr{label5}}
\addresses{
	\addr{label1} Ecole Nationale Sup\'{e}rieure de G\'{e}nie Energ\'{e}tique et Proc\'{e}d\'{e}s (ENSGEP),
	University of Abomey, \\Republic of Benin
	\addr{label2} Institute of Mathematics and Physical Sciences (IMSP), Republic of Benin
	\addr{label3} University of Abomey-Calavi, Department of Physics, Republic of Benin
	\addr{label4} GSMC, Faculty of Sciences and Techniques, Marien Ngouabi University, Brazzaville, Congo
	 and LMPHE, Faculty of Sciences Mohammed V University, Rabat, Morocco
	\addr{label5} Erciyes University, Department of Physics, 38039, Kayseri, Turkey
}

\date{Received April 2, 2019, in final form May 30, 2019}

\begin{document}

\maketitle

\begin{abstract}
We used  mean-field theory based on the Bogoliubov inequality for the Gibbs
free energy to examine the magnetic properties of a mixed spin-5/2 and spin-7/2 Blume-Capel ferrimagnetic system.
The thermal behaviours of the system
magnetization are classified according to the extended N\'eel nomenclature. The
system exhibits compensation phenomena where a complete cancellation of
sublattice magnetizations is observed below the critical temperature. Temperature-dependent phase diagrams are constructed for the case of unequal sublattice crystal
field interactions. Under appropriate conditions, our calculations reveal first-order
transitions in addition to second-order ones previously observed in Monte Carlo
simulations.

\pacs 61.10.Nz, 61.66.Fn, 75.40.Gb
\end{abstract}

\section{Introduction}
 Due to its theoretical interest and practical applications, the
 Blume-Capel (BC) model \cite{blume,capel} attracted intense attention in the last decades.
It has been successfully used to describe cooperative physical systems, in particular 
 multicomponent fluids, ternary alloys, $^3$He--$^4$He
 mixtures and various magnetic problems  \cite{lawrie}.
 The original model Hamiltonian comprises a single-ion
 anisotropy parameter and spin operators that can take the values $\pm 1,0$.
 The generalization of the BC model for spin values larger than~$1$ has
 been  investigated in detail 
 through different approaches  \cite{siqueira, siqueira1, 
chakraborty,kaneyoshi, kaneyoshi1, tucker, tucker1} (see 
   \cite{costabile} and references therein).
 Extension of the model to study two-sublattice
 mixed-spin systems with unequal magnetic moments or
 multiple layered structures of different magnetic substances
  has been a field of growing 
interest in recent years  \cite{svendsen,ozkan,karimou2018} due to their practical
 utilities such  as in thermomagnetic recording, electronic
 and magneto-optical readout devices  \cite{mollah, manriquez, jiang}.
 Indeed, many
 novel phenomena not observed in single-spin Ising systems were displayed,
 such as: giant magnetoresistance  \cite{binasch}, surface magnetic anisotropy  \cite{sayama}, etc.
 Under appropriate conditions, 
these systems exhibit a compensation temperature where the total magnetization
 vanishes below the critical temperature.
 The existence of compensation phenomena in ferrimagnets has
 a  technological significance  \cite{mollah, manriquez, jiang} since a small amount
 of driving field is needed
 to achieve  magnetic pole reversal or the sign change of the global magnetization.
At compensation points, the system cannot interact with external fields.
 The effect
 of the single-ion anisotropy (crystal field) strength 
 on critical and compensation
 properties has been theoretically studied by several authors
 using statistical-mechanical techniques: mean-field (MF)
 theory  \cite{kaneyoshi2,wang}, Monte Carlo
 (MC) simulations  \cite{wang1, jabar},
 effective-field (EF) theory  \cite{deviren, bouziane}, Bethe lattice approach with
exact recursion relations \cite{albayrak2010, eddahri}, etc., (see
 detailed references in  \cite{albayrak}). An obvious way to solve the 
two-dimensional mixed
 spin Ising model is to map it onto an exactly solved one. 
Such a method has been considered to solve
the mixed spin-1/2 and spin-$S$ ($ S > 1/2$) Ising models on
 the honeycomb lattice \cite{consalves}.
 Quite recently,
 Bahlagui et al. \cite{bahlagui, bahlagui1}
 addressed  the mixed spin 
$(5/2,7/2)$ Ising ferrimagnet on a square lattice by means of standard MC simulations. This mixed Ising
 system could describe several bimetallic molecular systems-based magnetic materials, in particular the GeFeO$_3$ as shown by the application of
 the Hund's rule to the Fe and Ge ions. The existence of compensation 
temperature in this system has been pointed out and its behaviour studied
 as a function of the strength of single-ion anisotropy.

In this work, using MF approximation, we studied the effects of two different crystal field
interactions on the magnetic behaviours of the same system investigated by Bahlagui et
al. \cite{bahlagui,bahlagui1}. Our calculations revealed some
 outstanding features of the system not previously
investigated as the temperature phase diagrams and the existence of first-order
transitions in the model. It is also revealed that the temperature dependence of the
global magnetization could be classified in the framework of the N\'eel nomenclature.
 The calculated temperature phase diagrams are illustrated in the plane
 of reduced temperature
versus sublattice crystal field.

The remainder part of this work is organized as follows. In section~\ref{sec2}, we
define the model and present its mean-field solution based on the Bogoliubov inequality
for the Gibbs free energy. In section~\ref{sec3}, we present and discuss the obtained results.
Finally, we conclude the present study in section~\ref{sec4}.

\section{The model and the mean-field solution} \label{sec2}
 We consider the mixed-spin Blume-Capel Ising ferrimagnetic system whose Hamiltonian includes
 the bilinear interaction $J$ between spins of the sublattices with spin-$5/2$ (sublattice A)
and spin-$7/2$ (sublattice~B) and the crystal field interaction
 constants $D_\text{A}$ and $D_\text{B}$ acting on the sublattice sites, respectively. This Hamiltonian
  is written as: 
       \begin{eqnarray}
     H =-J\sum_{\langle{i,j}\rangle}{S_{i}^\text{A}\sigma_{j}^\text{B}}
      - D_\text{A}\sum_{i}{(S_{i}^\text{A})^2}
      - D_\text{B}\sum_{j}{(\sigma_{j}^\text{B})^{2}},
      \label{1}
    \end{eqnarray}
where each spin $S_{i}$ located at site $i$ is a spin-$5/2$
 with six discrete spin values, i.e., $\pm5/2$, 
$\pm3/2$ and $\pm1/2$ and each spin 
$\sigma_{j}$ located at site $j$ is a spin-$7/2$ that can take on eight discrete
values  $\pm7/2$, $\pm5/2$, $\pm3/2$ and $\pm1/2$.

 The most direct way of deriving the mean-field equations
 is to use the variational principle for the Gibbs 
 free energy \cite{bogoliubov,feynmann}, 
  \begin{eqnarray}
    F(H) \leqslant \Phi \equiv F_{0}(H) + \langle  H - H_{0} \rangle _{0}\,,
   \end{eqnarray}
   where $F(H)$ is the true free energy of the model described by the Hamiltonian given in (\ref{1}). $F_{0}(H)$ is the average
   free energy calculated with a trial Hamiltonian $H_{0}$ 
which depends on variational parameters; $\langle  H - H_{0} \rangle _{0}$ denotes
   a thermal average of the value $H - H_{0} $ over the ensemble defined by the trial Hamiltonian~$H_{0}$.
   
   In this work, we use one of the simplest choices for this trial Hamiltonian which is given by:
   \begin{eqnarray}
    H_{0}=-\sum_{i}\big[\alpha_{S}S_{i}^\text{A}+ D_\text{A}(S_{i}^\text{A})^{2}\big] -\sum_{j}\big[\alpha_{\sigma}\sigma_{j}^\text{B}+ D_\text{B}(\sigma_{j}^\text{B})^{2}\big],
   \end{eqnarray}
  where $\alpha_{S}$ and $\alpha_{\sigma}$ are the two variational parameters related to the molecular
  field acting on the two different sublattices, respectively. Through this approach, we found the free energy
  and the equations of state (sublattice magnetizations per site $m_\text{A}$ and $m_\text{B}$) as follows:
   \begin{align}
   g&=\frac{\Phi}{N}=
   -\frac{1}{\beta}\ln\bigg[2\exp\left({\frac{25\beta D_\text{A}}{4}}\right)\cosh\left(\frac{5}{2}\beta\alpha_{S}\right) +
  2\exp\left({\frac{9\beta D_\text{A}}{4}}\right)\cosh\left(\frac{3}{2}\beta\alpha_{S}\right)\nonumber\\ 
  &+2\exp\left({\frac{\beta D_\text{A}}{4}}\right)\cosh\left(\frac{1}{2}\beta\alpha_{S}\right)\bigg]
   -\frac{1}{\beta}\ln\bigg[2\exp\left({\frac{49\beta D_\text{B}}{4}}\right)\cosh\left(\frac{7}{2}\beta\alpha_{\sigma}\right)\nonumber\\
   &+ 2\exp\left({\frac{25\beta D_\text{B}}{4}}\right)\cosh\left(\frac{5}{2}\beta\alpha_{\sigma}\right)\nonumber
  +  2\exp\left({\frac{9\beta D_\text{B}}{4}}\right)\cosh\left(\frac{3}{2}\beta\alpha_{\sigma}\right)\nonumber\\ 
  &+2\exp\left({\frac{\beta D_\text{B}}{4}}\right)\cosh\left(\frac{1}{2}\beta\alpha_{\sigma}\right)\bigg]
  +\alpha_{S}m_\text{A} +\alpha_{\sigma}m_\text{B}  -Jqm_\text{A}m_\text{B}\,,
  \label{4}
  \end{align}
where $\beta= \frac{1}{k_\text{B}T}$, $N$ is the total number of sites of the lattice and $q$ is the number
of the nearest neighbours of every spin
 of the lattice. The sublattice magnetizations
per site that appear in equation~(\ref{4}) are defined by:
\begin{align}
  m_\text{A}&=\frac{1}{2}\frac{5\sinh(\frac{5}{2}\beta\alpha_{S}) + 3\exp({-4\beta D_\text{A}})\sinh(\frac{3}{2}\beta\alpha_{S}) +
  \exp({-6\beta D_\text{A}})\sinh(\frac{1}{2}\beta\alpha_{S})}{\cosh(\frac{5}{2}\beta\alpha_{S}) +
  \exp({-4\beta D_\text{A}})\cosh(\frac{3}{2}\beta\alpha_{S}) +
  \exp({-6\beta D_\text{A}})\cosh(\frac{1}{2}\beta\alpha_{S})}\,, \label{5}\\
  m_\text{B}&=\frac{1}{2}\bigg[7\sinh\left(\frac{7}{2}\beta\alpha_{\sigma}\right) +   
  5\exp({-6\beta D_\text{B}})\sinh\left(\frac{5}{2}\beta\alpha_{\sigma}\right)+
  3\exp({-10\beta D_\text{B}})\sinh\left(\frac{3}{2}\beta\alpha_{\sigma}\right)\nonumber\\
  &+ \exp({-12\beta D_\text{B}})\sinh\left(\frac{1}{2}\beta\alpha_{\sigma}\right)\bigg]
  \times\bigg[\cosh\left(\frac{7}{2}\beta\alpha_{\sigma}\right) +
  \exp({-6\beta D_\text{B}})\cosh\left(\frac{5}{2}\beta\alpha_{\sigma}\right)\nonumber\\
  &+ \exp({-10\beta D_\text{B}})\cosh\left(\frac{3}{2}\beta\alpha_{\sigma}\right) +
  \exp({-12\beta D_\text{B}})\cosh\left(\frac{1}{2}\beta\alpha_{\sigma}\right)\bigg]^{-1}.
\end{align}
   
  Now, by minimizing the free energy in equation~(\ref{4}) with respect to $\alpha_{S}$ and $\alpha_{\sigma}$, we
  obtain
   \begin{eqnarray}
  \alpha_{S}=qJm_\text{B}\,, \qquad \alpha_{\sigma}=qJm_\text{A}.
  \label{7}
 \end{eqnarray}

The mean-field properties of the present model are given by equations~(\ref{4})--(\ref{7}). As the set of equations~(\ref{5})--(\ref{7}) have in general several solutions for the pair, the chosen pair is the one which minimizes the free
  energy.
  
In order to determine the compensation temperature, one should define the global magnetization $M_{T}$ of the model
which is given by:
\begin{eqnarray}
 M_{T}=\frac{m_\text{A}+m_\text{B}}{2}
\end{eqnarray}
and study its behaviour with the temperature.

\section{Numerical results and discussions} \label{sec3}
\begin{figure}[!t]
 \begin{center}
 \includegraphics[angle=0,width=0.6\textwidth]{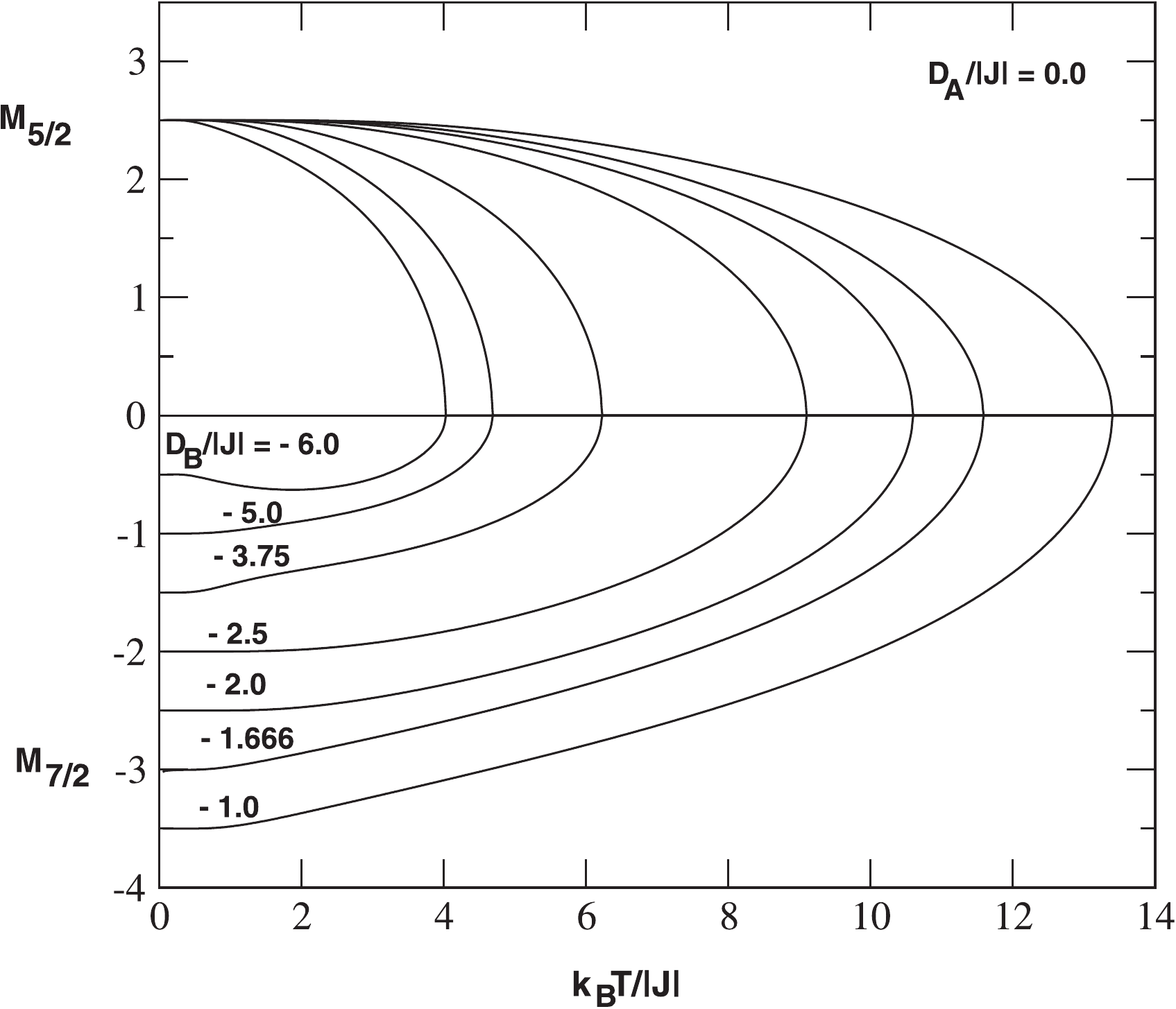}
 \end{center}
 \caption{Thermal variation of the sublattice magnetizations $M_{5/2}$ and $M_{7/2}$ for the Blume-Capel model defined in the text, 
 when the value of $D_\text{B}/|J|$ is varied for $D_\text{A}/|J|= 0$.}
 \label{fig1}
 \end{figure}

\begin{figure}[!t]
 \begin{center}
 \includegraphics[angle=0,width=0.6\textwidth]{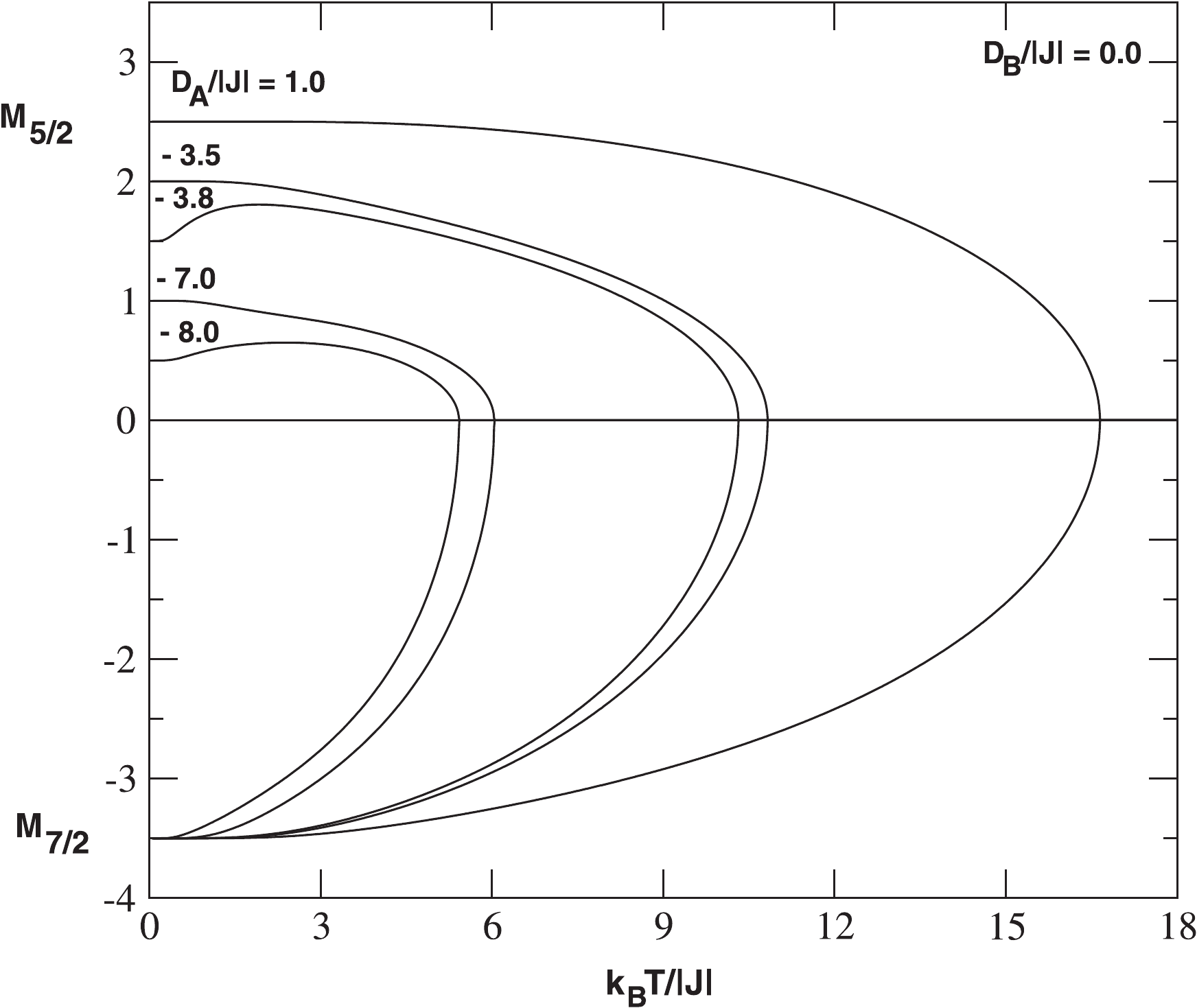}
 \end{center}
 \caption{Thermal variation of the sublattice magnetizations $M_{5/2}$ and $M_{7/2}$ for
 the mixed spin Blume-Capel model,
 when the value of $D_\text{A}/|J|$ is varied for $D_\text{B}/|J|= 0$.}
 \label{fig2}
 \end{figure}

In the previous section, we have derived expressions for the free energy,
sublattice and global magnetizations that would enable us to evaluate
 the transition temperatures for the mixed BC model. In this section, 
we show some results by solving them numerically.
  The computed data on the thermal variations
 of the order parameters which are the sublattice magnetizations 
are shown in figures~\ref{fig1} and  \ref{fig2} for selected values of 
 the reduced crystal field strengths
$D_\text{A}/|J|$ and $D_\text{B}/|J|$. Thermal variations of magnetizations are
 very useful in constructing thermal phase diagrams.
In figure~\ref{fig1}, $D_\text{A}/|J|$ is set to 0.
As expected, with an increasing temperature, $M_{5/2}$ falls
 from its unique saturation value $\pm5/2$, decreases monotonously and finally
vanishes. On the contrary, $M_{7/2}$ shows seven saturation values  with three hybrid ones:
 $-3$, $-2$, $-1$. Similar results have been reported in  \cite{ bahlagui, bahlagui1}.
 For an in-depth understanding of  these results, one should consider the problem
 from the energetic point of view and look for possible configurations at $T=0$.
 As displayed in figure~1 of  \cite{bahlagui}, for  $D_\text{A}/|J|=0$, one finds
 seven ground configurations comprising  three hybrid ones which are associated
 with coexistence lines of stable thermodynamic phases. 
 
In figure~\ref{fig2}, the effect of the crystal field $D_\text{A}/|J|$ is evaluated. Here, five saturation
 values are recovered for $M_{5/2}$ while $M_{7/2}$ shows a unique saturation value. 
 These results look similar to those displayed in figure~\ref{fig1}.
 Similar trends have been also reported in  \cite{bahlagui,bahlagui1}.
 
\begin{figure}[!t]
 \begin{center}
 \includegraphics[angle=0,width=0.58\textwidth]{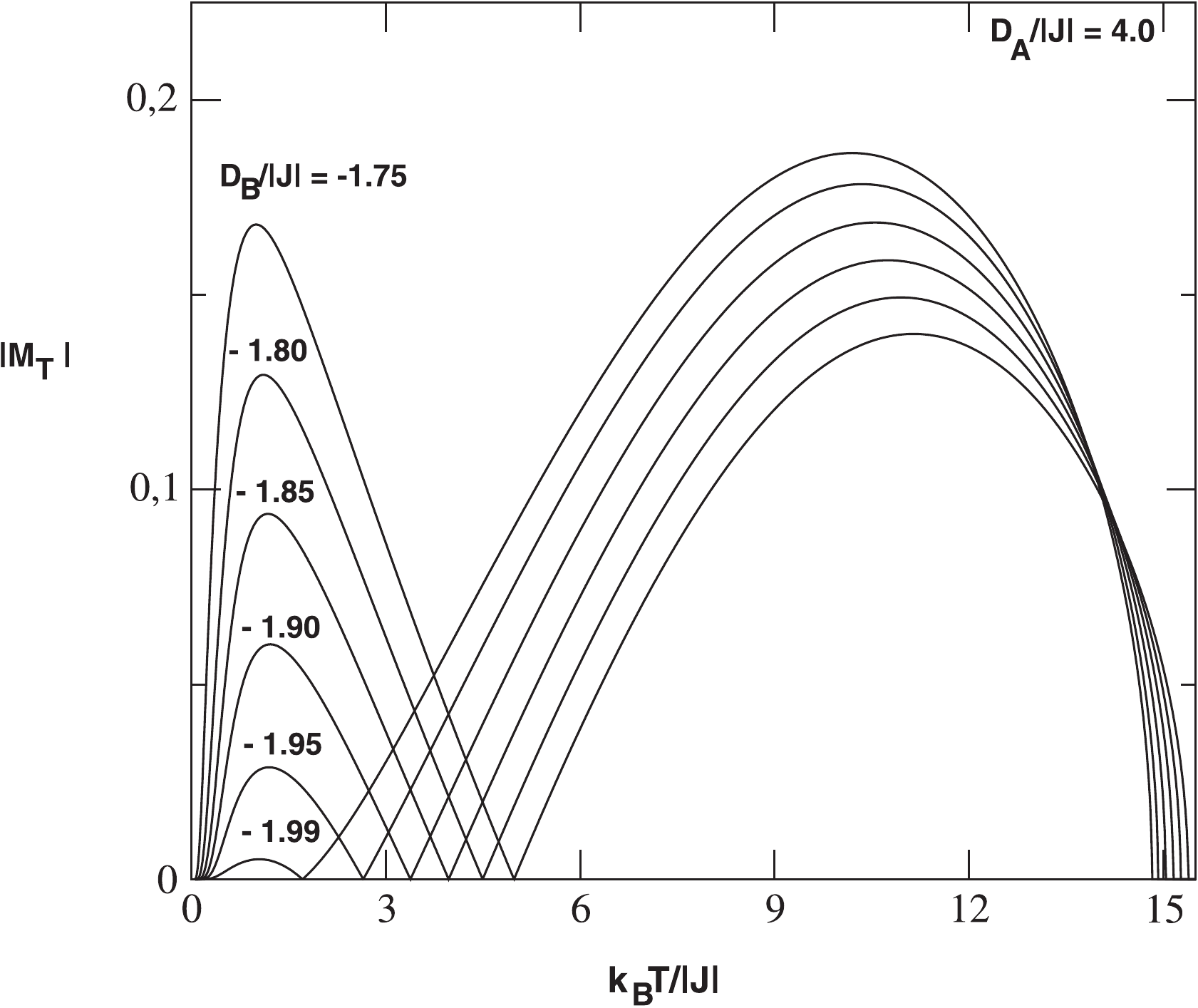}
 \end{center}
 \caption{Behaviour of the total magnetization of the system $|M_T|$ as a function of the 
temperature for $D_\text{A}/|J|= 4.0$ and negative values of $D_\text{B}/|J|$.}
 \label{fig3}
 \end{figure}

Figure~\ref{fig3} illustrates,  for $D_\text{A}/|J|=4$, the behaviour
 of the total magnetization per spin as a 
function of the temperature, for selected
values of the reduced crystal field $D_\text{B}/|J|$. It turns out that 
compensation phenomena exist in the model where the global magnetization vanishes below 
 the critical temperature. The strength of the crystal field 
affects both compensation and critical temperatures ($T_\text{comp}$ and $T_\text{c}$).
 An increase of the absolute value of this field leads to a decrease  
of $T_\text{comp}$. This behaviour also agrees with results from  \cite{ekiz,espriela}.

\begin{figure}[!t]
 \begin{center}
 \includegraphics[angle=0,width=0.6\textwidth]{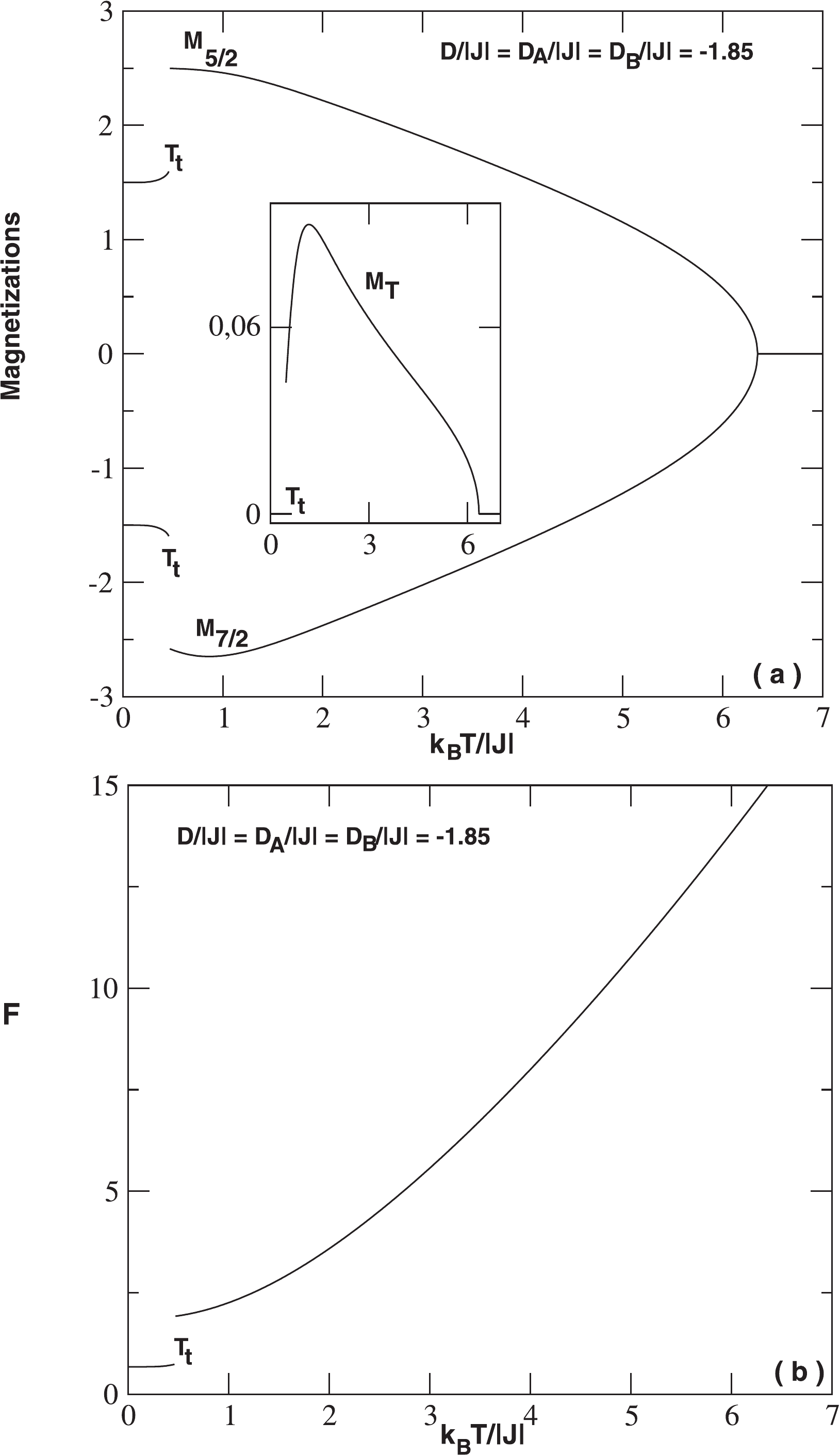}
 \end{center}
 \caption{Temperature dependence of sublattice magnetizations, the total magnetization and the free energy
  for $D/|J|= D_\text{A}/|J|= D_\text{B}/|J|= -1.85$. From different panels, one can conclude that the model exhibits first-order transitions
  where jumps appear in different thermodynamical quantities presented. 
 $T_\text{t}$ indicates the first-order temperature.}
 \label{fig4}
 \end{figure}

In figure~\ref{fig4},   the thermal
 variations of sublattice and global magnetizations and the free energy of the system for 
$D_\text{A}/|J|=D_\text{B}/|J|=D/|J| - 1.85$ are illustrated. An interesting trend is observed at 
low temperature: the existence of simultaneous jumps in four physical  quantities.
This is a strong proof of the existence of first-order
 transitions in the model for specific values of the model parameters.
 Such results are certainly the most exciting 
ones generated
 through  the present study. The first-order transition temperature
 is indicated in figure~\ref{fig4} by $T_\text{t}$.
 
\begin{figure}[!t]
 \begin{center}
 \includegraphics[angle=0,width=0.75\textwidth]{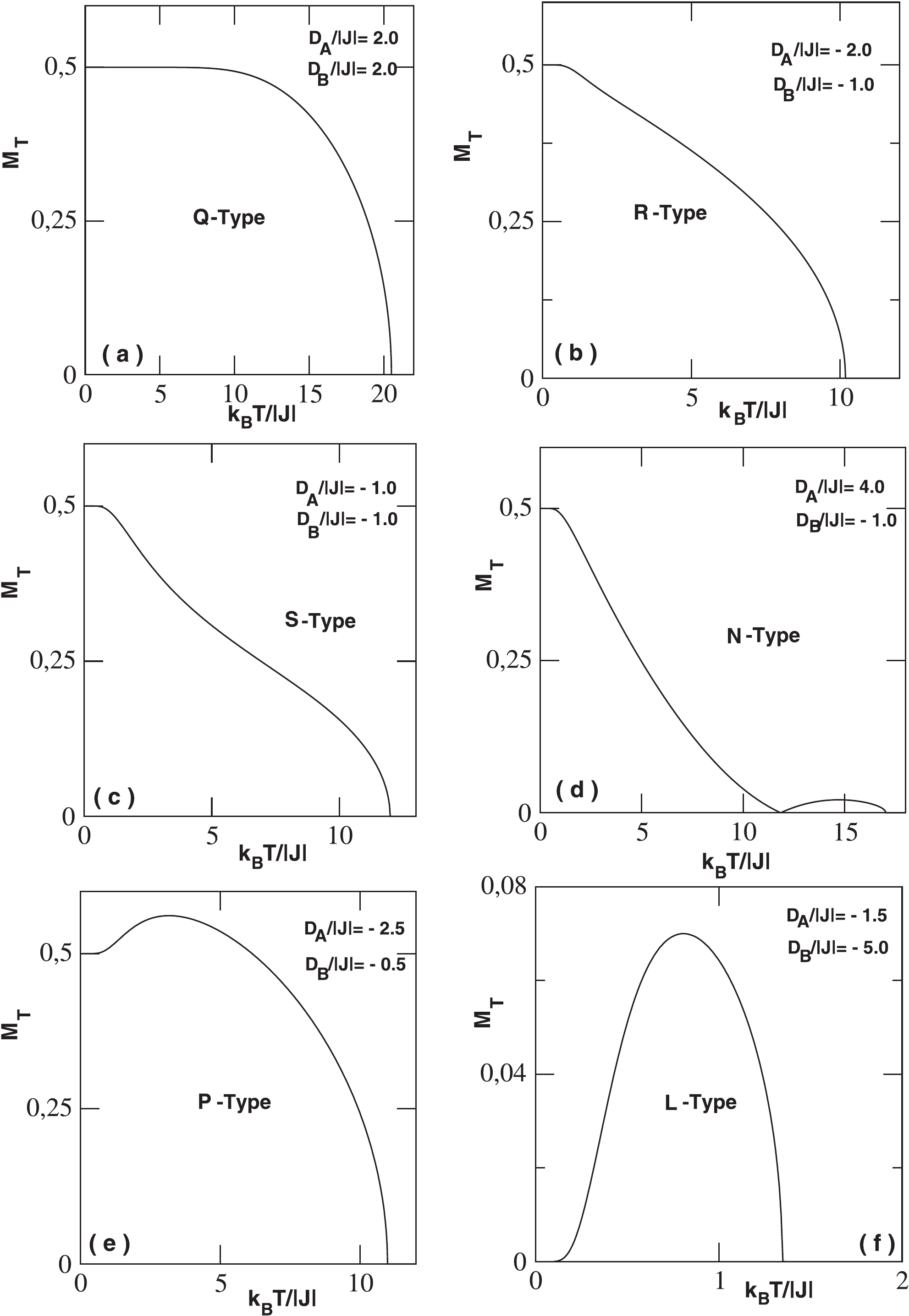}
 \end{center}
 \caption{Temperature-dependence of the total magnetization $M_T$  for selected values of the 
crystal field. The model shows the 
  Q-, R-, S-, N-, P- and L-types of compensation behaviours as classified in the extended N\'{e}el nomenclature.}
 \label{fig5}
 \end{figure}

 Figure~\ref{fig5} indicates the temperature-dependence of the global
magnetization of the system. Obviously, six types of behaviours, namely 
Q-, R-, S-, N-, P- and L-types,  are obtained according to their classification
 in the extended  N\'{e}el nomenclature \cite{a10,a11,a12}. 

The previous results on the thermal variations of the order parameters are useful
 in drawing the phase diagrams of the system. We 
 present and illustrate these findings on the phase 
diagrams in the $(D_\text{A}/|J|, k_\text{B}T/|J|)$ plane
for constant values of $D_\text{B}/|J|$ and on the $(D_\text{B}/|J|, k_\text{B}T/|J|)$ plane
for constant values of $D_\text{A}/|J|$. The case of equal 
strengths of the reduced crystal fields, $D_\text{A} = D_\text{B}$ on
the  $(D/|J|, k_\text{B}T/|J|)$ plane was also investigated.
 In  different phase diagrams constructed,
 the solid, dotted and dashed lines, 
respectively, denote second-order transition,
 first-order transition and compensation lines.

\begin{figure}[!t]
 \begin{center}
 \includegraphics[angle=0,width=0.7\textwidth]{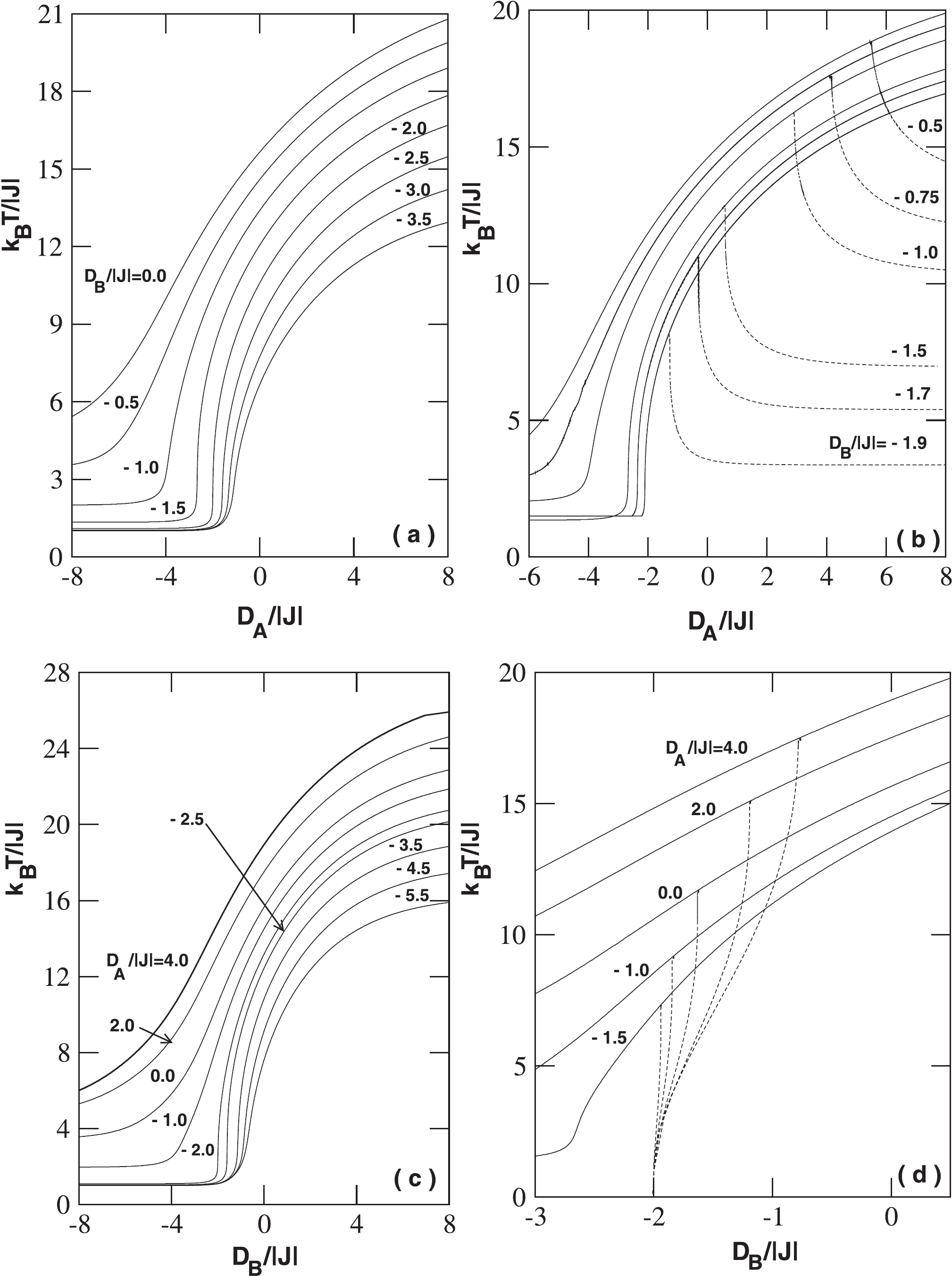}
 \end{center}
 \caption{While the phase diagrams on the $( D_\text{A}/|J|, k_\text{B}T/|J|)$ plane for constant values of $D_\text{B}/|J|$ are given in (a) and (b), 
  the roles of the crystal field are illustrated in (c) and (d). (a) and (c) only present the second-order transition lines
  which are labeled with $D_\text{B}/|J|$ and  $D_\text{A}/|J|$ values, respectively. (b) and (d) show the second-order transition lines
  and the lines of the compensation temperatures.}
 \label{fig6}
 \end{figure}

Through figure~\ref{fig6}, it emerges that the critical temperature $T_\text{c}$, for not
 too large values of the sublattice crystal field decreases with decreasing  values
of the field strength. On the contrary, 
 for relatively large values of the sublattice crystal field, 
it appears that the critical temperature becomes not sensitive to the
 strength of that field. Similar results have been reported in reference
  \cite{dakhama} where a new approach is used to generate
 the exact phase diagrams
 of the mixed spin-1/2 and spin-$S$ Ising model on a square lattice. Therein, 
the derived exact critical temperature tends to the exact value of the 
 nearest neighbour spin-1/2 Ising model obtained by Onsager \cite{onsager} as 
$D/J\rightarrow -\infty$.  In figure~\ref{fig6}~(a),  for $D_\text{A}/|J|\rightarrow
\infty$ $(-\infty)$ and   $D_\text{B}/|J|\rightarrow -\infty$, the second-order
 transition temperature reaches a constant value at $12.93$ $(1.02)$
and for  $D_\text{B}/|J|\rightarrow \infty$, this value of $k_\text{B}T_\text{c}/|J|$ is at 
$20.79$ $(5.43)$. Similarly, in figure~\ref{fig6}~(c),
as $D_\text{B}/|J|\rightarrow \infty$ $(-\infty)$ for  $D_\text{A}/|J|\rightarrow -\infty$,
 the second-order transition temperature also
reaches a constant value at $11.41$ $(1.02)$
and for  $D_\text{A}/|J|\rightarrow \infty$, the constant value of $k_\text{B}T_\text{c}/|J|$ is
 at $25.91$ $(6.01)$. In figure~\ref{fig6}~(b),  values of  $D_\text{B}/|J|$ are used
 to label transition lines. 
The second-order transition  and  compensation temperature lines
 are displayed from top to bottom for  $D_\text{B}/|J|= -0.5, -0.75,
-1.0, -1.5, -1.7$ and $- 1.9$. In figure~\ref{fig6}~(d), the roles of the sublattice
 crystal field were interchanged with respect to figure~\ref{fig6}~(b).
The figure illustrates  lines of compensation
 temperature that merge at about  $D_\text{B}/|J|= -2.0$ and terminate there.
 
The constant value of $k_\text{B}T_\text{c}/|J|$ for negative values of $D$ originates from the 
fact that the lattice spins have the values $\pm 1/2$ in this region,
 the system being in the 
antiferromagnetic $(+1/2, -1/2)$ phase. The value $1.02$ found is about twice
 the value $0.56$ that we derived from MC calculations (not presented herein) on a system of size
 $N=100$ using $10^5$ MC steps per site. The MC result is of the order of the value
 $4k_\text{B}T_\text{c}/|J|=2.104$ found in reference  \cite{kane} from an exact formulation of the
spin-5/2 BC model. For large positive values of $D$, the system is in the phase
 $(5/2,-7/2)$ and the limiting value for the critical temperature is again about
 twice the value $11.3$ calculated in the above-mentioned reference.

\begin{figure}[!t]
 \begin{center}
 \includegraphics[angle=0,width=0.55\textwidth]{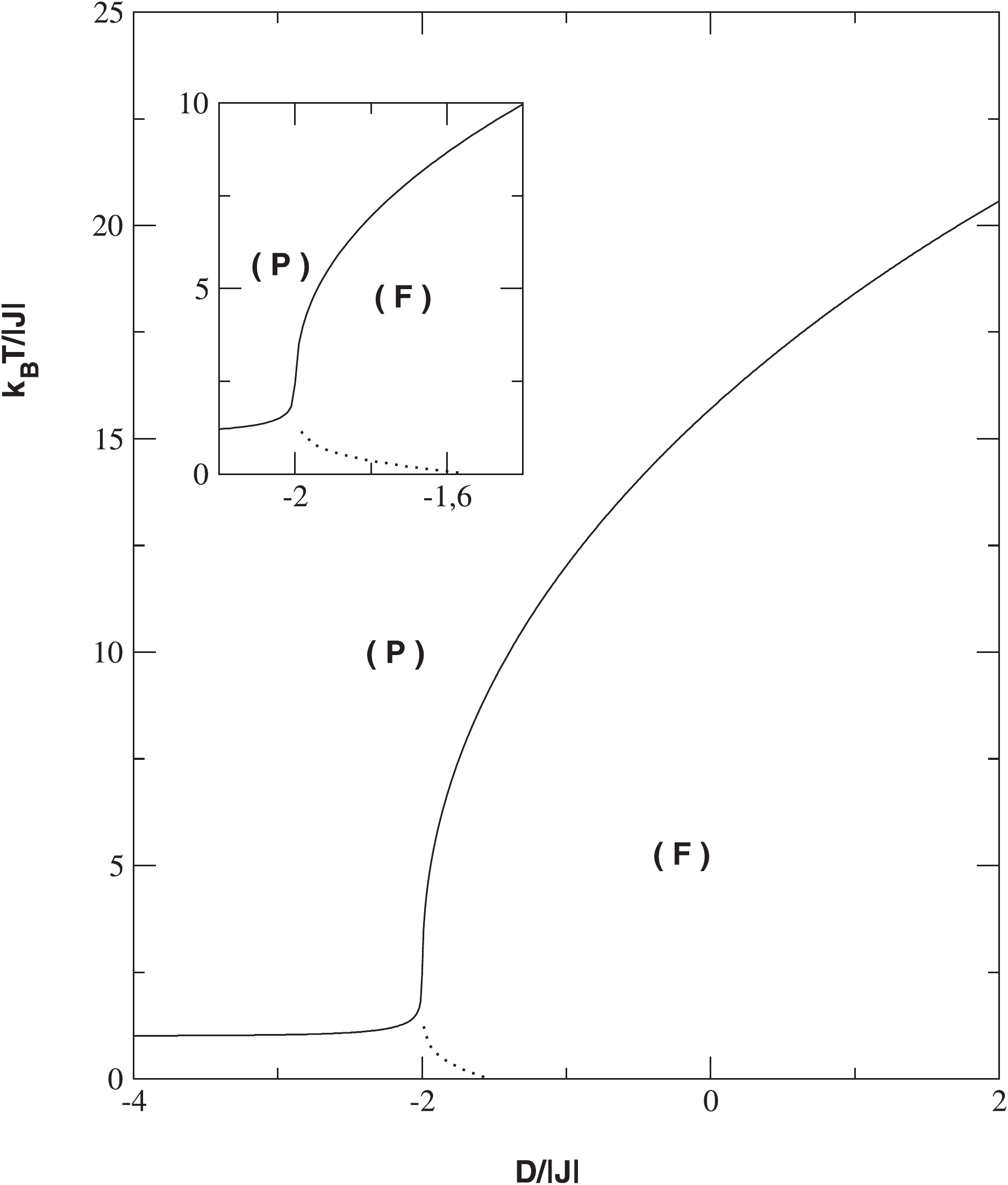}
 \end{center}
\caption{ The phase diagram for equal values of the sublattice crystal fields  $D/|J|= D_\text{A}/|J|= D_\text{B}/|J|$, in the  $( D/|J|, k_\text{B}T/|J|)$ plane.
  The solid and dotted lines indicate the second- and first-order transition lines.}
 \label{fig7}
 \end{figure}

 In figure~\ref{fig7}, we have depicted the phase diagram for the case of equal strengths of 
sublattice crystal fields. It turns out from the figure that
 when the value of the reduced crystal field  $D/|J|\rightarrow \infty$
$(-\infty)$, the second-order phase transition temperature reaches the
 limiting values at about $25.22$ $(1.02)$. 
  It is important to mention that figure~\ref{fig7} presents some resemblances  with
 figure~10 of \cite{karimou2016}. Moreover, the system exhibits
 a first-order transition at low temperature and negative values of the crystal field.
 The corresponding transition line ends at $D/|J|=-2.0$ and does not connect to 
the second-order line to generate a tricritical point. 
 This line presents some resemblances with the first-order 
  transition line illustrated in figure~1 of~\cite{albayrak3}.

\newpage

\section{Conclusion} \label{sec4}
  In this work, we examined the magnetic and critical properties of 
the  mixed spin-$5/2$ and spin-$7/2$ Blume-Capel Ising
  system using the mean-field theory. The thermal variations of the order parameters
 have been shown. The behaviour of the
global magnetization has been elucidated using the extended
 N\'{e}el classification nomenclature. Thus,
  Q-, R-, S-, N-, P- and L-types of compensation behaviours are got for
 appropriate values of the system parameters. Our results are
 in qualitative agreement with those computed by Monte Carlo simulations
 in reference  \cite{bahlagui}. In particular our figures~\ref{fig1}; \ref{fig2}; \ref{fig3} bear
resemblances respectively with figures 6 and 7; 2 and 4; 9 of that reference.
  More interesting is the 
existence of the first-order transitions that we detected in the low 
 temperature regime while
 analyzing the system behaviour
 for the case of equal crystal field strengths for 
 both sublattices. 
Second-order phase transitions are also present with the existence
 of compensation points. Corresponding lines in the model parameters' space
have been presented.

\vspace{-5mm}
\ukrainianpart
\title{Фазові діаграми і критична поведінка змішаної  спін-5/2 і спін-7/2 системи Ізінга}
\author{M. Каріму\refaddr{label1,label2}, Р.А. Єссуфу\refaddr{label2,label3}, Г. Дімітрі Нганцо\refaddr{label4}, Ф. Гонтінфінд\refaddr{label2,label3}, E. Албайрак\refaddr{label5}}
\addresses{
	\addr{label1} Національна вища школа енергетики та процесів науки  (ENSGEP), університет м. Абомей,\\ Республіка Бенін
	\addr{label2} Інститут математики і фізичних наук (IMSP), Республіка Бенін
	\addr{label3} Університет м. Абомей-Калаві, фізичний відділ, Республіка Бенін
	\addr{label4} GSMC, факультет природничих наук і технологій, університет Маріана Нгуабі, Браззавіль, Конго і LMPHE, факультет природничих наук, університет  V Мохаммеда, Рабат, Морокко	
	\addr{label5} Університет Ерджієс, фізичний відділ, 38039, Кайсері, Туреччина
}

\makeukrtitle

\begin{abstract}
Ми застосували середньопольову теорію, що базується на нерівності Боголюбова для вільної енергії Гіббса, для  вивчення магнітних властивостей змішаної  спін-5/2 і спін-7/2 феромагнітної системи Блюма-Капела. Температурна поведінка намагніченості системи класифікується відповідно до розширеної номенклатури Нееля. Система демонструє компенсаційне явище, де спостерігається повне скасування намагніченості підграток нижче критичної температури. Для випадку неоднакових взаємодій кристалічного поля підграток побудовано температурно залежні фазові діаграми. 
При певних умовах наші обчислення виявили перехід першого роду додатково до переходу другого роду, що  був спостережений раніше в симуляціях Монте Карло.

\keywords середньопольова теорія, спін-змішана система, намагніченість, компенсаційна температура, фазові переходи
\end{abstract}

\end{document}